% last modified by Juan Aug 29 .
% Revised on Mar 2 2003 by Juan

\input harvmac.tex
\input epsf.tex
%\draft

\def\+{^\dagger}
\def\d{d}

\def \bi{\bibitem}

\def \lr { \lref}
\def\np {{  Nucl. Phys. }}
\def \pl {{  Phys. Lett. }}
\def \mpl {{ Mod. Phys. Lett. }}
\def \prl {{  Phys. Rev. Lett. }}
\def \pr  {{ Phys. Rev. }}

\def \bi{Born-Infeld}

%--------+---------+---------+---------+---------+---------+---------+
%Title page

\Title{
 \vbox{\baselineskip10pt
  \hbox{hep-th/9708147}
  \hbox{PUPT-1718}
  \hbox{RU-97-68}
 }
}
{
 \vbox{
\centerline{Brane Dynamics}
\centerline{From the Born-Infeld Action}
 }
}
\vskip -25 true pt

\centerline{
 Curtis~G.~Callan,~Jr.\footnote{$^1$}{e-mail:  callan@viper.princeton.edu}}
\centerline{\it Joseph Henry Laboratories,
Princeton University, Princeton, NJ  08544}
\centerline{ and}
\centerline{J.M.~Maldacena\footnote{$^2$}{e-mail:malda@physics.rutgers.edu} }
\centerline{{\it Physics Department, Rutgers University,
Piscataway NJ 08855}}

\centerline {\bf Abstract}
\smallskip
\baselineskip10pt
\noindent

We use the abelian \bi\ action for the worldvolume gauge field and
transverse displacement scalars to explore new aspects of
D-brane structure and dynamics. We study several classic gauge field
configurations, including
point charges in any worldvolume dimension and vortices in two
worldvolume dimensions, and show that, with an appropriate excitation
of the transverse coordinate field, they are BPS-saturated solutions.
The Coulomb point charge solutions turn out to represent, with considerable
fidelity, fundamental strings attached to the brane (their magnetic
counterparts describe D1-branes attached to D3-branes).  We also
show that S-matrix for
small excitations propagating on the point charge solution
is consistent with (and gives further illuminating
information about) Polchinski's effective open string boundary condition.

\Date {August 1997}

\noblackbox
\baselineskip 14pt plus 1pt minus 1pt

%--------+---------+---------+---------+---------+---------+---------+
%Bibliography

\lref\mast{J.M.~Maldacena and A.~Strominger, Rutgers preprint RU-96-78,
hep-th/9609026.}
\lref\mst{J.M.~Maldacena and A.~Strominger, \prl 77 (1996) 428,
hep-th/960.}

\lref\gkgrey{S.S.~Gubser and I.R.~Klebanov, Princeton preprint
PUPT-1648, hep-th/9609076.}

\lref\kr{B.~Kol and A.~Rajaraman, Stanford preprint SU-ITP-96-38,
SLAC-PUB-7262, hep-th/9608126.}

\lref\gkk{G.~Gibbons, R.~Kallosh and B.~Kol, Stanford preprint
SU-ITP-96-35, hep-th/9607108.}

\lref\fks{S.~Ferrara, R.~Kallosh, and A.~Strominger, Phys.~Rev.~D~{\bf 52}
(1995) 5412, hep-th/9508072.}

\lref\hmf{{\it Handbook of Mathematical Functions},
M.~Abramowitz and I.A.~Stegun, eds. (US Government Printing Office,
Washington, DC, 1964) 538ff.}

\lref\GR{I.S.~Gradshteyn and I.M.~Ryzhik, {\it Table of Integrals,
Series, and Products}, Fifth Edition, A.~Jeffrey, ed. (Academic Press:
San Diego, 1994).}

\lref\Unruh{W.G.~Unruh, Phys.~Rev.~{D}14 (1976) 3251.}

\lref\ascv{A.~Strominger and C.~Vafa,
Phys. Lett.  B379 (1996) 99, hep-th/9601029.}

\lref\cama{C.G.~Callan and J.M.~Maldacena,
Nucl. Phys.  B472 (1996) 591, hep-th/9602043.}

\lref\ms{J.M.~Maldacena and L.~Susskind, Stanford preprint
SU-ITP-96-12, hep-th/9604042.}

\lref\dmw{A.~Dhar, G.~Mandal and S.~R.~Wadia,
Tata preprint
TIFR-TH-96/26, hep-th/9605234.}

\lref\dm{S.R.~Das and S.D.~Mathur,
 hep-th/9606185; hep-th/9607049.}

\lref\dmI{S.R.~Das and S.D.~Mathur,  hep-th/9601152.}

\lref\hms{G.~Horowitz, J.~Maldacena and A.~Strominger, hep-th/9603109.}

\lref\HM{G.~Horowitz  and A.~Strominger, \prl  77 (1996) 2368,
hep-th/9602051.}

\lref\gkt{J.P.~Gauntlett, D.~Kastor and J.~Traschen,
hep-th/9604179.}

\lref\us{S.S.~Gubser and I.R.~Klebanov, hep-th/9608108.}

\lref\dgm{S.~Das, G.~Gibbons and S.~Mathur, hep-th/9609052.}

\lref\ktI{I.R.~Klebanov and A.A.~Tseytlin, Princeton
preprint PUPT-1639, hep-th/9607107.}

\lref\KT{I.R. Klebanov and A.A. Tseytlin, \np B475 (1996) 179,
hep-th/9604166.}

\lref\at{ A.A. Tseytlin, \np B475 (1996) 149, hep-th/9604035.}

\lref\ATT{A.A. Tseytlin, \mpl A11 (1996) 689,  hep-th/9601177.}

\lref\CY{M. Cveti\v c and D. Youm, \pr D53 (1996) 584, hep-th/9507090;
Contribution to `Strings 95',  hep-th/9508058. }

\lref\CT{M. Cveti\v c and  A.A.  Tseytlin, hep-th/9606033.}

\lref\CTT{M. Cveti\v c and  A.A.  Tseytlin, \pr D53 (1996) 5619,
 hep-th/9512031.}

\lref\CYY{M. Cveti\v c and D. Youm, hep-th/9603100.}

\lref\juan{J. Maldacena, hep-th/9605016.}

\lref\kaaa{ R. Kallosh, A. Linde, T. Ort\'in, A. Peet and A.  Van
Proeyen, Phys. Rev. D{46} (1992) 5278.}

\lref\myers{C. Johnson, R. Khuri and R. Myers, hep-th/9603061,
Phys. Lett. B378 (1996) 78.}

\lref\gibb{G. Gibbons, Nucl. Phys. {B207} (1982)  337;
P. Breitenlohner, D. Maison and G. Gibbons, Commun. Math. Phys.
{ 120} (1988) 295.}

\lref\HS{G.T. Horowitz  and A. Strominger, \prl 77 (1996) 2368,
hep-th/9602051.}

\lref\fkk{S. Ferrara and  R. Kallosh,
hep-th/9602136;
hep-th/9603090;
S. Ferrara, R. Kallosh, A. Strominger, Phys. Rev. D{52} (1995)
5412, hep-th/9508072. }

\lref\LWW {F. Larsen and F. Wilczek,  \pl B375 (1996) 37,
hep-th/9511064; hep-th/9609084.}

\lref\age{D. Page, \pr D13 (1976) 198.}

\lref\maha{J. Maharana and J.H. Schwarz, \np B390 (1993) 3,
hep-th/9207016.}

\lr \brek{J.C. Breckenridge, R.C. Myers, A.W. Peet and C. Vafa,
hep-th/9602065;
J.C. Breckenridge, D.A. Lowe, R.C. Myers,
 A.W. Peet, A. Strominger and C. Vafa,
\pl B381 (1996) 423, hep-th/9603078.}

\lr \pertuu{
G. Gilbert,  hep-th/9108012;
 C.F.E. Holzhey  and F. Wilczek, Nucl. Phys. B380 (1992) 447,
 hep-th/9202014; R. Gregory and  R. Laflamme,  Phys. Rev. D51 (1995) 305,
 hep-th/9410050;
 Nucl. Phys. B428 (1994) 399,
 hep-th/9404071.}

\lr \khuri{R. Khuri, \np B376 (1992) 350.}

\lr \schw{J.H. Schwarz, \np B226 (1983) 269.    }
\lr\bho{E. Bergshoeff, C. Hull and T. Ort\'in, \np B451 (1995) 547, hep-th/9504081.}

\lref\dmI{S.R.~Das and S.D.~Mathur, Phys. Lett. B375 (1996) 103,
 hep-th/9601152.}

\lref\sendsix{ A. Sen, {\it Strong Coupling Dynamics of Branes
from M-theory}, hep-th/9708002.}

\lref\bankssusskind{T. Banks and L. Susskind,
{\it Brane - Anti-brane Forces},
 hep-th/9511194.}

\lref\andyending{ A. Strominger,
 Phys .Lett .B383 (1996) 44,
 hep-th/9512059.}
\lref\townsendending{ P. Townsend,
Phys. Lett. B373 (1996) 68,
hep-th/9512062.}

\lref\polchinski{ J. Polchinski, {\it Tasi Lectures on D-branes},
hep-th/9611050.}

\lref\leigh{R. Leigh,
Mod. Phys. Lett .A4 (1989) 2767.}

\lref\schwarz{ M. Aganagic, C. Popescu and J. Schwarz,
Nucl. Phys. B495 (1997) 99,
 hep-th/9612080.
}

\lref\diaconescu{ D. Diaconescu, {\it D-branes, Monopoles
and Nahm Equations},  hep-th/9608163.}

\lref\wsinstanton{ X. Wen and E. Witten, Phys. Lett.
166B (1986) 397.}

\lref\andyhol{ K. Becker, M. Becker and A. Strominger,
Nucl. Phys. B456 (1995) 130,
hep-th/9507158.}

\lref\vafahol{C. Vafa,
 M. Bershadsky, C. Vafa and  V. Sadov,
Nucl. Phys. B463 (1996) 420, hep-th/9511222;
 Nucl. Phys. B463 (1996) 398,
 hep-th/9510225.}

\lref\wheeler{ I've seen this in the book Gravitation, Do you know
a better reference?. This is JAW's charge without charge proposal
and it is certainly described in MTW. There is certainly a paper of
Wheeler, perhaps with a collaborator where the idea first appears.
I imagine the ref is in MTW. }

\lref\juanandy{ J. Maldacena and A. Strominger,
unpublished notes.}

\lref\dls{ M. Douglas, D. Lowe and J. Schwarz,
private comunication.}

\lref\wittenntwo{ E. Witten, {\it Solutions of four dimensional
field theories via M-theory}, hep-th/9703166.}

\lref\aharony{
O. Aharony, J. Sonnenschein and S. Yankielowicz,
ucl. Phys. B474 (1996) 309, hep-th/9603009.}

\lref\gibbons{Lecture at Centro de Estudios Cientificos de Santiago,
August 1997.}

\lref\mdgm{ M. Douglas and G. Moore, {\it D-branes, quivers and
ALE instantons}, hep-th/9603167.}

\lref\dgauge{M. Douglas, {\it Gauge fields and D-branes},
hep-th/9604198.}

%--------+---------+---------+---------+---------+---------+---------+
\newsec{Introduction}

The \bi\ action for D-brane
dynamics is a  $p+1$-dimensional $U(1)$ gauge theory with $9-p $ neutral
scalars describing transverse fluctuations of the brane. This action is
the dimensional reduction of $10$-dimensional supersymmetric \bi\
electrodynamics \refs{\polchinski,\leigh}.
For a variety of reasons, most treatments of this system have been based on
a linear, Maxwellian, approximation to this nonlinear action. In this
paper we will take the specific nonlinearities of the BI action seriously,
revealing some interesting new insights into the statics and the dynamics
of intersecting branes. On the static side, we will show that fundamental
strings attached to branes are described by Coulomb point charge solutions of
\bi\ electrodynamics. More elaborate brane intersections
correspond to other localized gauge objects such as the magnetic
monopole and the vortex line. Some of the essential features of the
energetics of these solutions come from supersymmetry and BPS arguments,
of course, but the dynamics rely on the specific nonlinear properties of the
action. This is particularly true of the S-matrix describing the way
small fluctuations on one brane can \lq leak through\rq\ to the other
brane. This S-matrix determines the effective boundary condition
one of the branes provides for the worldsheet field theory on the other.
Its study provides interesting insights into Polchinski's open string theory
definition of D-branes.

In what follows, we will explain these results in detail, commenting as we
go on their reliability and their implications for other aspects of
string theory and M-theory.
%--------+---------+---------+---------+---------+---------+---------+

\newsec{Strings from $U(1)$ point charges: A Linear Argument}

We begin by using BPS arguments to find some solutions of the p-brane
worldsheet gauge theory which have the interpretation of strings ending on
branes and  D-p branes  ending on D-(p+2)-branes. Let us first assume that the
massless excitations of a p-brane are described by the dimensional
reduction of the $10$-dimensional  Maxwell action. This ignores important
nonlinearities (which we will consider later) but will provide a useful
orientation. The supersymmetry variation of the gaugino is
\eqn\gauginovar{\delta \chi = \Gamma^{\mu\nu} F_{\mu\nu} \epsilon}
where $\mu\nu$ are ten-dimensional indices. A BPS background is one
where $\delta\chi=0$ for some $\epsilon$. Now imagine we have a point charge
Coulomb gauge field, or $A_0 = c_p/r^{p-2} $ where $r$ is the spatial
$p$-dimensional distance and $c_p$ is fixed by some charge quantization
condition. So far only $F_{0r}\ne 0$ and \gauginovar\ shows that there are
no preserved supersymmetries  since  $\Gamma^{0r}$ has no zero eigenvalues.
We can, however, also excite one of the transverse scalar fields. Indeed,
$X^9 = c_p/r^{p-2}$ is a solution of the linearized field equation for $X^9$
for which $F_{9r}\equiv \partial_r X^9 = F_{0r}$. In the
presence of both excitations \gauginovar\ reduces to
\eqn\preservedsusy{( \Gamma^{0r} + \Gamma^{9r}) \epsilon = 0~~\Rightarrow~~
(\Gamma^{0} + \Gamma^{9}) \epsilon = 0.}
The usual arguments tell us that half of all choices for $\epsilon$
satisfy this condition. In other words, this background preserves
half the supersymmetries originally possessed by the p-brane and is
still a BPS-saturated state.

What is the interpretation of such a state? The excitation of the $X^9$ field
that we have chosen amounts to giving the brane a transverse \lq spike\rq\
protruding in the $\hat 9$ direction and running off to infinity.
It turns out that this spike must be interpreted as a fundamental
string attached to the D-p-brane. To show this, we calculate the energy of the
configuration. Of course, the energy of a Coulomb field is strictly infinite,
but let us look at this infinity in more detail by calculating the energy for
$r \ge \delta $ where $\delta $ is small. We find\foot{We are
setting $\alpha' =1$ and using  conventions in which $g \rightarrow 1/g$
under S-duality.}
\eqn\energy{
E(\delta) = { 1\over (2 \pi)^p g} \int  d^p r  {1\over 2} \{ F_{0r}^2 +
  F_{09}^2  \} = {c_p^2 (p-2)^2 \over  (2 \pi)^p g}
 \int d^p r {1\over r^{2(p-1)}} =
 {c_p^2 (p-2) \Omega_{p-1}  \over  (2 \pi)^p g}  { 1\over \delta^{p-2}}~,
}
where $\Omega_p$ is the volume of the unit $p$-sphere.
Remarkably, this energy is proportional to the value of $X^9$ at $\delta$:
\eqn\energydelta{
E(\delta) = { c_p  \Omega_{p-1} ( p-2 ) \over (2 \pi)^p g} X^9(\delta)~.
}
In the limit $\delta\to 0$ the transverse \lq spike\rq\ is infinitely
long and has infinite energy by virtue of having a constant energy per unit
length. This interprets the Coulomb energy divergence as being due to
an infinitely long string attached to the brane! We should be able to tell
just what kind of string it is from the value of its energy per unit length.

Since we expect the source of a worldvolume {\it electric}
field to be the fundamental string, we should be able to see that \energy\
matches the energy per unit length of a fundamental string. To do this,
we need to identify the charge quantization constant $c_p$. Consider first
a 1-brane with (necessarily constant) gauge field $F_{01}$ excited. Depending
on the strength of $F_{01}$, this is a type IIB D-string with
 some number $n$ of adsorbed fundamental strings. From the BPS mass formula
we know that the tension of such excited strings is $(1/g + g n^2/2)T_f $ for
small $n$ and string coupling $g$ (where  $T_f = 1/2\pi$ is  the
fundamental string tension).  In the linearized weak field limit of the
\bi\ action the additional tension due to the electric field is
$(1/2 g) F_{01}^2 T_f $.
This implies that the quantization condition on the one-dimensional electric
field is $F_{01} = g n $. The higher-dimensional cases are related to this by
T-dualities and one can see that the general quantization condition on
the electric flux is
\eqn\qcond{
{ 1 \over (2 \pi)^{p-1} } \int_{S_{p-1}} F_{0r}  = g n~.}
This implies that $c_p (p-2) \Omega_{p-1}/(2\pi)^{p-1} = g $
 and that the energy of the unit charge Coulomb solution is
\eqn\energyfinal{
E(\delta) = { 1\over  2 \pi} X^9(\delta) = T_f X^9(\delta)
}
This is the energy of a fundamental string of length $X^9(\delta)$.
The analysis works for any value of $p$ and tells
us that there is a BPS gauge configuration describing the attachment of a
fundamental string to {\it any} D-brane. This is exactly what Polchinski's
string theoretic description of D-brane solitons would lead us to expect
\polchinski .

There is one further possibility for solutions of the kind we are discussing.
On a $3+1$-dimensional worldvolume (D3-brane) there exist magnetic point
charge solutions for which $F_{\hat \theta \hat \phi} = N c_m/ r$ (and $c_m$
will be determined by magnetic monopole charge quantization conditions)\foot{
Were $\hat \theta,~\hat \phi$ denote unit vectors.}. Taking the
transverse coordinate excitation to be $X^9 = N c_m/r$, the condition for
unbroken susy now reads
\eqn\susydstring{
( \Gamma^{\hat \theta \hat \phi} + \Gamma^{9r} ) ~\epsilon =0\qquad{\rm or}\qquad
\Gamma^{1239} ~\epsilon = \epsilon ~.
}
This is exactly what we expect for the preserved susy when a type IIB D1-brane
along $\hat 9$ is attached to a D3-brane along $123$. The energy analysis of
this solution is consistent with this picture and shows that $c_m =\pi$,
independent of the string coupling $g$. Among other things, this implies that
the shape of the solution is independent of the string coupling.
Clearly the same solution would describe a D-$p$-brane ending on a
D-$(p+2)$-brane.

\vskip 1cm
\vbox{
{\centerline{\epsfxsize=2.5in \epsfbox{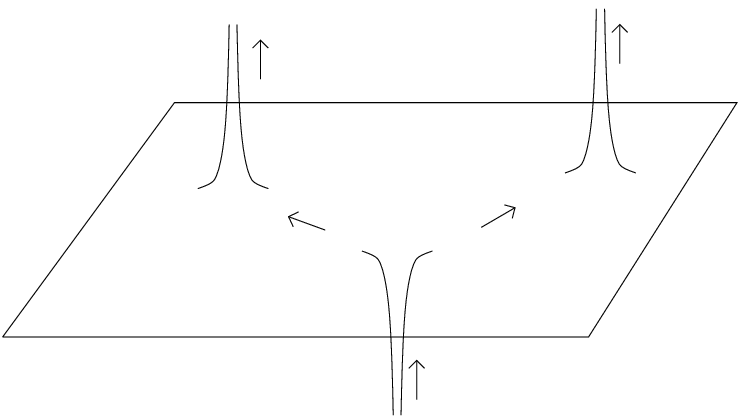}}}
{\centerline{\tenrm FIGURE 1: Branes with
multiple strings attached on both sides.
}}
{\centerline{ The arrows represent the direction
of the electric field,
}}
{\centerline{which also defines the orientation of the
string}}
}
\vskip .5cm

It is interesting that in any of the above solutions we may replace the
single-center Coulomb solution with a completely general solution of
Laplace's equation:
\eqn\supperp{
 1/r^{p-2} \rightarrow \sum  \pm { 1 \over | \vec r- \vec r_i |^{p-2} }
}
where for each $i$ we have a choice of signs. When this function is used to
specify $X^9$, we see that it represents multiple strings ending at arbitrary
locations on a brane (see Figure 1). The two signs of the charge correspond to
strings attached to different sides of the brane, but having the same
orientation. We could say that strings with positive signs in \supperp\
are ending on the brane while the ones with negative sign are `departing'
from the brane. If we had  strings of different orientations the solution
would not be BPS. A careful analysis
shows that the energy is independent of the locations of the attachment points
\eqn\energypar{ E = \sum_i | X^9(\delta_i)| T_f }
and is exactly sum of the lengths of the various strings times the
fundamental string mass per unit length. An important point is that there
is no extra constant term to be identified as an `attachment energy'.

Note that this type of solution
allows us to describe the deformation of a fundamental string traversing
the brane into two open strings ending on the brane at separate points.
It might seem paradoxical that two solitons with opposite charges are
not atracted to each other. What happens is that they have also a
scalar charge (associated to $X^9$) and two opposite scalar charges
repel each other. In other words: the Coulomb energy  is exactly
balanced by the elastic deformation energy of the brane.
In short, we now have a BPS description of the breaking of a fundamental
closed string into two open strings when it meets a D-brane. Notice that the
same description applies to a D1-brane meeting a D3-brane.

There are implications for M-theory to be drawn from this as well.
Consider a D4 brane intersecting fundamental type IIA strings as above. This
should be the same as the $X^{11}$ compactification of the M-theory membrane
ending on an M-theory fivebrane along  a line. Since the IIA
D-fourbrane action is the dimensional reduction of the M-fivebrane action
we can `lift' the solutions we have found to solutions of the
M-fivebrane action that are invariant
along $X^{11}$. This is indeed the solution describing an M-twobrane
ending on a fivebrane and the intersection is one dimensional and is oriented
along $X^{11}$. After the usual redefinition of parameters, the
quantization conditions  work out exactly right. From what we now know
about how the D4-brane can \lq slice\rq\ fundamental strings, we learn that when
an M-theory membrane intersects a fivebrane on a line, the membrane can
break into {\it open} membranes attached to the fivebrane, which can then
separate and move independently \refs{\andyending,\townsendending}.
This motion is a deformation respecting the BPS condition, a property which
has been used to calculate the entropy of four-dimensional black holes \mst .

%--------+---------+---------+---------+---------+---------+---------+
\newsec{Strings from point charges: Non-Linear Treatment}

In the previous section we made use of the properties of the quadratic,
Maxwellian, limit of the action for the worldsheet degrees of freedom
of D-branes. Insofar as we are interested in the properties of strict
BPS states (state counting and masses), we probably can't go wrong in doing
so. However, as soon as we interest ourselves in non-BPS states, no matter how
close in energy to the BPS limit, the Maxwellian action gives, as we shall see,
qualitatively incorrect results. Sooner or later, therefore, we will be obliged
to redo the calculations of the previous section using the fully
nonlinear BI action. That will be the topic of this section.

Pure \bi\ $U(1)$ electromagnetism in ten dimensions is described by the action
\eqn\biten{L_{10} = -{1\over( 2 \pi )^9  g}\int d^{10}x ~\sqrt{det(1+F)}}
where $F_{\mu\nu}$ is the  gauge field strength. To reduce this
to the dynamics of a $p$-brane, identify $A_i$ for $i=0,1,\ldots, p$ with
the worldbrane gauge field and $A_a$ for $a=p+1,\ldots, 9$ with the
transverse displacements of the brane (and only allow fields to depend
on the worldbrane coordinates $x^i$) \polchinski . We won't attempt a completely
general analysis of this system, but will focus on one or two instructive
special cases.

First, consider the case where the worldbrane gauge field
is purely electric and only one transverse coordinate (call it $X$) is
excited. In that case, the worldbrane action reduces to
\eqn\electact{L=-{1\over  g_p}\int d^p x
        \sqrt{(1-\vec E^2)(1+\vec\nabla X^2)+
(\vec E\cdot\vec\nabla X)^2 - \dot X^2 }~.}
The normalization corresponds to the background tension of a D-$p$-brane in
its ground state (hence the factor of $1/g_p$,  where we have defined $g_p
\equiv g (2 \pi)^p $). Since we are going to want
the energy function anyway, it is best to pass to the Hamiltonian formalism
right away. The canonical momenta associated with $\vec A_i$ and $X$,
respectively, are
\eqn\canmom{\eqalign{
 g_p\vec\Pi&=
{\vec E(1+\vec\nabla X^2)-\vec\nabla X(\vec E\cdot\vec\nabla X)\over
\sqrt{(1-\vec E^2)(1+\vec\nabla X^2)+(\vec E\cdot\vec\nabla X)^2-\dot X^2}}\cr
 g_p P &={\dot X\over
\sqrt{(1-\vec E^2)(1+\vec\nabla X^2)+(\vec E\cdot\vec\nabla X)^2-\dot X^2}}}}
and, with a little algebra, one can construct the Hamiltonian:
\eqn\biham{ H={1\over   g_p}\int d^p x
\sqrt{(1+\vec\nabla X^2)(1+g^2_p P^2)+
     g^2_p\vec\Pi^2+g^2_p(\vec\Pi\cdot\vec\nabla X)^2}~.}
The canonical momentum $\vec\Pi$ is of course subject to the constraint
$\vec\nabla\cdot\vec\Pi=0$ (this follows from the absence of a canonical
momentum for $A_0$) and we can use this to turn \biham\ into an action
for the transverse displacement alone. (Since we assumed that the magnetic
field vanishes, we have to be careful not to look for solutions that violate
this condition.)

Our primary goal is to reproduce the configurations described in the previous
section, so we shall first look for static solutions. We first solve the
constraint by setting
\eqn\consolv{
g_p \vec\Pi = \vec\nabla\Lambda, \qquad  \nabla^2 \Lambda = 0 ~\Rightarrow~
      \Lambda = \Sigma_i{q_i\over |\vec r-\vec r_i|^{p-2}}   }
where the charges $q_i$ are the same as the ones in the
last section. Since the charge quantization condition is a long range
property,  far from the center we can apply the
usual weak field charge quantization conditions.
The static equation for $X$, gotten by varying
\biham\ after setting $P=0$ and constraining $\Pi$, is
\eqn\xequatn{
\vec\nabla\cdot \bigl[ {\vec\nabla X+g_p^2\vec\Pi
(\vec\Pi\cdot\vec\nabla X)\over
 \sqrt{1+\vec\nabla X^2+g^2_p\vec\Pi^2+
     g^2_p(\vec\Pi\cdot\vec\nabla X)^2}}\bigr]=0~.
}
We now observe that if
we set $\vec\nabla X=g_p\vec\Pi$, \xequatn\ reduces to the identity
$\vec\nabla\cdot\vec\Pi=\nabla^2\Lambda=0$ and therefore defines a solution
(in which $X=\Lambda$). Substituting the solution in \biham\ to get the energy,
we find that the expression under the square root becomes a perfect square and
the energy simplifies:
\eqn\bpsen{E = {1\over  g_p}\int d^p x (1 + \vec\nabla\Lambda^2)~.}
Note how the complicated energy function has \lq linearized\rq\ and
the  coefficient of the quadratic term ($1$ instead of $1/2$)
correctly reflects the sum of equal linearized energy contributions of the $X$
field and the gauge field. The net result is that the \bi\ theory has
{\it exactly} the same BPS solutions (and not just the same energies)
as the linearized theory. This is a
rather stronger result than we are strictly entitled to expect.
Since these solutions saturate the BPS energy bound (the energy is the
string tension times the length of the strings) we expect them to
be annihilated by the appropriate supersymmetries. It should be possible
to check this explicitly from the formulas in \schwarz\ for the
supersymmetry variations of the full \bi\ action, but we have yet to
check this point.

There is also an interesting set of static solutions representing
strings going between branes and antibranes.
Branes and antibranes attract each other but for small coupling they
will move towards each other very slowly so that our solutions make
sense for some long time.
The simplest way to find them  is to exploit spherical symmetry by
choosing a single-charge electric solution of the constraint and
looking for the most general spherically symmetric solution of \xequatn:
\eqn\nonbps{
g_p \vec\Pi = {A\over r^{p-1}}\hat r \qquad
  {X^\prime(1+A^2/r^{2p-2})\over
 \sqrt{(1+A^2/r^{2p-2})(1+{X^\prime}^2)}}={B\over r^{p-1}}.}
The constant of integration $A$, since it is associated with a conserved
momentum, is a conserved quantity and is related to the fundamental string
tension because we are studying the electric solution (to be precise,
$A = (p-2) N c_p$). The other constant of integration, $B$, is arbitrary and
varying it changes the physics of the solution.

The equation for $X$ can easily be integrated to give
\eqn\xsoln{ X(r) = \int_r^\infty dr {B\over\sqrt{r^{2p-2}-r_0^{2p-2}}}}
where we have defined a new length parameter $r_0^{2p-2}=B^2-A^2$
(well-defined for $B>A$). This solution only makes sense
for $r>r_0$ since $X^\prime(r_0)=\infty$, even though $X(r_0)$ is
finite (see Figure 2b). There is a natural continuation of the
surface through $r=r_0$, back out through increasing $r$, given by
\eqn\xcont{ \tilde X(r) = 2 X(r_0) - \int_r^\infty dr
{B\over\sqrt{r^{2p-2}-r_0^{2p-2}}}~.}
This joins smoothly onto \xsoln\ at $r=r_0$ but adopts the other sign of the
square root for $X^\prime$.  The result is two parallel p-branes,
joined by a throat of finite length and thickness, separated asymptotically by
a distance $\Delta \equiv 2 X(r_0)$ (see Figure 2c).
The fact that one is a brane and the other an antibrane can be seen by
defining an orientation on the first brane, then moving
continuously to the other brane and seeing that the orientation changes.

\vskip 1cm
\vbox{
{\centerline{\epsfxsize=5in \epsfbox{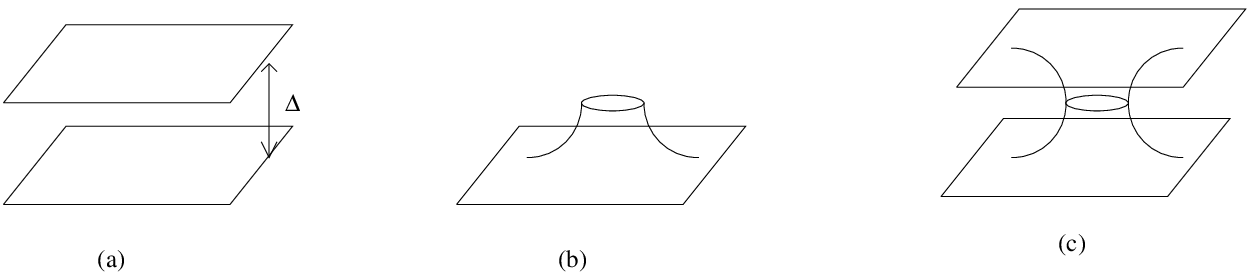}}}
{\centerline{\tenrm FIGURE 2:}}
{\centerline{ (a) A brane-antibrane separated by a distance
$\delta$,
%{\centerline{
 (b) Throat solution described by \xsoln ,}}
{\centerline{ (c) Full solution  by patching together
two solutions  as in (b).}}
}
\vskip .5cm

Fixing the distance between the branes $\Delta$ and the number of strings
(which we do by regarding the charge parameter A as fixed), we have an
equation to solve to fix the parameter $r_0$ in the solution. There are
two solutions which, for large $\Delta$, are approximately
\eqn\twosol{
r_0 \sim  A/\Delta ,~~~~~~~~~~~ r_0 \sim \Delta
}
up to a numerical constant (see Figure 3).
In the first case, the radius of the throat goes to zero as
$\Delta \to \infty$, a situation we interpret as a fundamental string running
between the brane and the antibrane. The second solution consists of a
large throat joining the branes and is easily seen to be a local
maximum of the action. Note that for the
second solution (Figure 3 b) the electric field is a small perturbation and
could be turned off without changing the solution dramatically.
In the limit that the electric field is zero, it is tempting to
interpret this second solution as a sphaleron associated to decay
of the brane-antibrane system. Given the existence of a sphaleron, there
ought to exist a `bounce' solution describing the decay of a brane-antibrane
system by tunneling under the barrier whose height is defined by the sphaleron.
A bounce is a solution of the Euclidean classical equations of motion for the
brane having a topology like that shown in Figure 3b (where Euclidean time
would be one of the horizontal directions in the figure). Such solutions can
indeed be found and describe decay of a brane-antibrane pair by nucleation
of a throat joining the brane pair (followed by an expansion of the throat
which converts brane worldvolume to kinetic energy of the throat expansion
front).  Although this is a possible picture of the decay, it is not the
dominant decay mode: while the tunneling rate is of order $e^{-1/ g}$,
the brane and antibrane also attract each other via long-range forces
which cause them to collide and annihilate by direct string processes
on a time scale of order ${1/ \sqrt{g}}$. Since this is much smaller than
the timescale of the throat nucleation process, it seems that the tunneling
is never physically meaningful\foot{We would like to thank Gary Horowitz for
pointing out an error in our treatment of this point in an earlier version
of this paper.}.
One could imagine trying to take some limit of the theory where the
attraction of the branes goes to zero  while the tension remains finite.
Such limits do not exist for branes of dimension $p \ge 1$. This is
of course consistent with the fact that it has not been possible to
quantize branes of dimension $p\ge 2 $, if it were possible to take
all distance scales to zero while keeping the brane tension finite,
then we would have a situation where the branes would be fundamental.

\vskip 1cm
\vbox{
{\centerline{\epsfxsize=5in \epsfbox{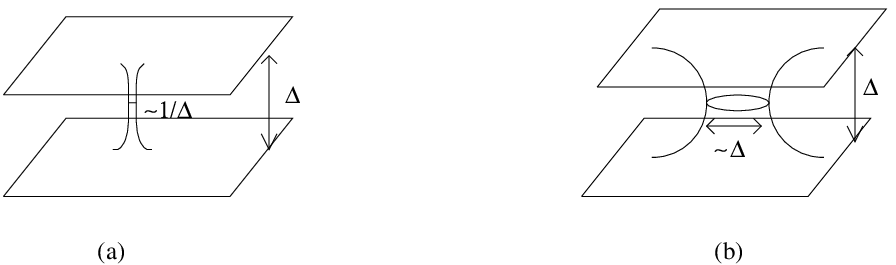}}}
{\centerline{\tenrm FIGURE 3: Brane - Anti-brane configuration
with a throat between them.}}
{\centerline{ (a) Solution corresponding to a
fundamental  string stretched in between.}}
{\centerline{
(b) Unstable configuration. }}
}
\vskip .5cm

Note that one doesn't find solutions of this type for
parallel branes (as opposed to brane-antibrane). We know that parallel branes
at small separation should correspond to monopoles of a spontaneously
broken non-abelian worldsheet gauge theory \diaconescu. We have not
understood the exact relation between the abelian description of parallel
brane-antibrane pairs and the nonabelian description of their brane-brane
analog.

Although this solution makes use of the full nonlinear \bi\ action and
is not a BPS state, it is
amusing to see how close it comes to satisfying the linear BPS condition
\gauginovar. The essential issue is the comparison of the electric field
$E=F_{0r}$ with the gradient of the scalar field,
$F_{9r}=X^\prime$. From
 the expression of the canonical momentum and the detailed form of the
solution we read off that
\eqn\brokensusy{ E = {A \over \sqrt{r^{2p-2}-r_0^{2p-2}}} =
{A\over B} X^\prime~.}
Substituting this result into \gauginovar, we find that the requirement
for an unbroken susy is
\eqn\badsusy{ (\Gamma^0 +\alpha \Gamma^9)\epsilon = 0 \qquad
          \alpha = A/B~.}
Unless $A/B=1$, this is not consistent and the susy is completely
broken. As we take the limit $r_0\to 0$, the  brane pair moves further
and further apart and $A/B\to 1$ (for the first solution in \twosol ).
In the limit, supersymmetry and stability are recovered.

At this point, we want to say a few words about limits of validity of
the solutions of \bi\ we have been discussing. If we replace $c_p$ by
$ N c_p$ in the point charge solutions, we get a solution describing $N$
superposed strings ending on a single D-brane. Since $c_p\sim g$ in the
electric (fundamental string) case and $c_p\sim 1$ in the magnetic (D-string)
case, we really have two parameters, $N$ and $g$, to vary in exploring the
validity of our solutions. The most important region is the `throat',
roughly defined as $r\sim X^9$, where
the transition from one type of p-brane to another takes place.
Strict validity of the \bi\ action requires not so much that the field
strengths $F_{\mu\nu}$ be small, but rather that their derivatives (measured
in units of $\alpha^\prime$) be small. A bit of arithmetic on the \bi\ point
charge solution shows that it is valid (in the sense just described)
when $N c_p\gg 1 $. Actually, to validate the neglect of gravitational
effects (we have been considering branes in flat space) we also need
$g^2 N \ll  1$ in the electric case and $gN\ll 1$ in the magnetic case.
As far as the throat region goes, this is all possible for large $N$ and
small $g$ so long as $g\gg 1/N\gg g^2$ (electric case) or $1\gg 1/N\gg g$
(magnetic case). Inside the throat region, toward $r=0$,
fields and field gradients become large and an application of the above
arguments would seem to indicate that the \bi\ solution can't be trusted
in that regime. And yet, we have seen that a naive application of the \bi\
equations reproduces the correct string tension and we will soon
see that more subtle features of the string dynamics are also reproduced.
It seems obvious that much of what \bi\ tells us about the singular region
is correct and that the question of limits of validity of solutions is
fairly subtle. For the purposes of this paper, therefore, we will take an
uncritical attitude toward this question, just to see how far we get.

%--------+---------+---------+---------+---------+---------+---------+

\newsec{S-Matrices and Boundary Conditions}

We have established that the \bi\ action has static solutions which correspond
to strings terminating on D-branes (and also to certain D-branes terminating
on higher-dimensional D-objects).  We
have to look at the behavior of small fluctuations propagating on our
static solutions with special attention to how they reflect from (or
transmit through) the effective junction between two kinds of brane.
As we shall see, it is here that the specific nonlinearities of the action
become quite crucial.

We take the BPS solution for a string attached to a 3-brane as background and,
for simplicity, study the propagation of perturbations of a scalar
field describing motion  in a direction
perpendicular to {\it both} branes. We denote the fluctuation coordinate by
$\eta$ and look at s-wave fluctuations in the obvious spherical
coordinate system. The linearized small fluctuation expansion of the \bi\
action about the exact background turns out to be
\eqn\fluct{-(1+{g^2c^2\over r^4})\partial_t^2\eta
     +r^{-2}\partial_r(r^2\partial_r\eta) = 0~.}

The entire effect of the nonlinearities is contained in the $r^{-4}$ term
multiplying the $\partial_t^2$: apart from that, the equation is
just the linear $3+1$-dimensional D'Alembertian. For future interpretation,
recall that the field giving the static transverse displacement of the
3-brane is $X^9=gc/r$. In the limit $r\to\infty$, the extra term can
be neglected, and one has the usual interpretation of spherical waves
freely propagating in three dimensions. One might have thought that the
$r^{-4}$ term just acts as a potential from which the otherwise free waves
scatter. In fact, it is so singular that the usual interpretation does not work
and $r\to 0$ functions as a second asymptotic regime to which waves can
escape. To see this, drop the constant term in the coefficient of
$\partial_t^2\eta$ in \fluct\ and make the change of variables $w=gc/r=X^9(r)$.
The result is very simple:
\eqn\rtozero{(-\partial_t^2 +\partial_w^2)\eta = 0 ~,}
the equation for free propagation in one dimension (at the correct
velocity of light)! This is the correct dynamical counterpart to the
identification of the static background solution as a string attached
to a threebrane. It relies in an essential way on the specific nonlinear
structure of the \bi\ action and, of course, explores almost, but not exactly,
BPS physics.

In order to fully verify the correspondence with Polchinki's
picture of D-brane dynamics via open fundamental strings, we have
to look into the effective boundary condition for small
fluctuations on the string (the $r\to 0$ region) imposed by the
presence of the three-brane (the $r\to\infty$ region). In other
words, we want the S-matrix connecting the two asymptotic regions.
We can get most of the information we want via the following
sequence of arguments. We first reduce to a purely radial problem
by projecting \fluct\ on frequency $\omega$:
\eqn\reduce{ \bigl(
{d^2\over dx^2}+1+{\kappa^2\over x^4}\bigr)\eta=0 \qquad x=\omega
w = { \omega g c \over r}  \qquad  \kappa = gc\omega^2~.
}
 This shows that the problem has
an interesting joint dependence on string coupling, string tension
and energy. Then we make a coordinate change suggested by WKB
considerations\foot{We want to thank K.~Saviddi for assistance on
this point.}: \eqn\nucoord{ \xi(x) =\int_{\sqrt{\kappa}}^x dy
\sqrt{1+{\kappa^2\over y^4}}\qquad \eta = (1+{\kappa^2\over
x^4})^{-1/4} \tilde \eta ~.} This transforms \reduce\ to
\eqn\nueqn{ \bigl(-{d^2\over d\xi^2}+ V(\xi)\bigr)\tilde\eta =
\tilde\eta \qquad V(\xi) = {5\kappa^2 \over (x^2+\kappa^2/x^2)^3}
} where $\xi$ now runs over the full real line while the potential
$V(\xi)$ vanishes as $\xi\to\pm\infty$ and is non-zero, but
finite, in the region of $\xi\sim 0$. Consequently, in the new
variables, we have a perfectly conventional one-dimensional
scattering problem.

A closer look at how the potential scales shows that, in various limits,
we may make the replacements
\eqn\potlim{ \eqalign{
V(\xi) \sim &{a\over\sqrt{\kappa}} \delta(\xi) \qquad \kappa\to 0\cr
V(\xi) \sim  &0  \qquad \kappa\to\infty }}
where $a$ is a dimensionless constant. We are interested in the behavior
of the one-dimensional reflection and transmission amplitudes $R$ and $T$.
Simple considerations about delta-function potentials show that
\eqn\smatrix{\eqalign{
R\to -1\qquad T\to & 2i\sqrt{\kappa}/a \qquad {\rm as}~~\kappa\to 0\cr
R\to 0\qquad T\to &1\qquad\qquad {\rm as}~~\kappa\to\infty
}}
and we could, if we wanted, learn more about the energy variation of these
amplitudes.

We can now make contact with the Polchinski picture of D-branes as boundaries
for open strings. Since $\tilde\eta$ is a disturbance on the string
attached to the threebrane, transverse to both the brane and the
string, it should, on Polchinski's picture, satisfy a Dirichlet (fixed)
boundary condition. In terms of our one-dimensional scattering theory picture,
this amounts to requiring $R=-1$ (and $T=0$). Our analysis of the scattering
problem shows that this is indeed the low-energy limit of the S-matrix but
that, as the energy increases, the S-matrix goes over from perfectly
reflecting to perfectly absorbing. The energy scaling of \reduce\ indicates
that the characteristic energy of the transition between the two behaviors
is $\omega_{crit}\sim 1/\sqrt{gc}$. The parameter $c$ is related to the string
tension and we find that, for the fundamental (electric) string case,
$\omega_{crit}\sim 1/\sqrt{g \alpha^\prime}$.
While, for the D-string (magnetic)
case $\omega_{crit}\sim 1/\sqrt{\alpha^\prime}$. In the weak-coupling $g\to 0$
limit, the Dirichlet  boundary condition accurately describes the scattering
physics of the fundamental open string at large energies compared to
the string
scale and therefore presumably captures the essential string theory dynamics.
On the other hand, in the magnetic or D-string case, the Dirichlet  boundary
condition only describes the physics below the string scale and thus only
captures the field theory limit of the problem.

\newsec{ Higher dimensional solitons}

In previous sections we considered excitations of the branes which
included only one transverse field. Now we consider excitations
involving two transverse fields (transverse positions). While
the solutions in this section are well known
\refs{\wsinstanton,\vafahol,\andyhol,\ascv}, we believe that we will
present them under a new and interesting light. We also want to
constrast them with our previous discussion.

Consider a two brane. In fact one can consider any relativistic brane
(a D-brane, M-brane, or even a fundamental string in Euclidean space)
but just concentrate on two of its spatial coordinates.
Denote by $z = X^1 + i X^2 $  two of the worldvolume coordinates
 and let  $ w = X^3 + i X^4 $ be two of the transverse coordinates.
All the nontrivial dynamics will involve only the $(z,w)$ variables.
The linear equation for fluctuations of $w$ is just the laplacian
$\partial_z \partial_{\bar z } w =0$. A solution to this equation
is to take $w(z,\bar z)$ to be holomorphic (i.e. $w(z)$). A crucial point
is that a holomorphic function is a solution of the full Nambu action
for
the brane and, in the supersymmetric case, these holomorphic solutions
are also BPS \refs{\wsinstanton,\andyhol}.
This can be recognized as the usual condition for worldsheet
instantons for the case of fundamental strings.
One interesting solution  is $w = c /z$, where $c$ is a constant.
We see that as we approach $z \rightarrow 0 $ we sweep out the
$ w$ plane, so near the origin the surface is really another brane stretched
along the $w$ plane. A pair of D2 branes intersecting at a point
is represented by $zw =0$ so that the deformation $zw = c$ is the single smooth
surface you get by blowing up the intersection point between the original pair
of D2 branes into a smooth finite throat. In the D-brane case (reinstating the
full \bi\ action) there can also be a gauge field. Any pure gauge field
(holomorphic vector potential and vanishing field strength) is in principle
acceptable, but the configuration $A_z = A_z^\infty + a/z - A_w^\infty c/z^2 $
is the only one for which $A_z$ stays finite as $z \to \infty$ and
the transformed field $A_w = { \partial z \over \partial w } A_z=
A_w^\infty - a/w - A_z^\infty c/w^2 $ stays finite as $w\to\infty$.
The constants $A_z^\infty$, $A_w^\infty$ are the Wilson lines that would
be present on two non-intersecting D2 branes. The extra parameter, $a$,
is a new Wilson line or modulus which is only allowed when the branes
intersect. Since $a$ is complex, there are two new real moduli.
One of them is measured by integrating the vector potential along
a contour which circles the origin of the $z$ plane and the other
is measured by a contour which goes through the throat and emerges in the $w$
plane (this last one would be finite if the branes were finite).
Overall, there are four extra real
parameters characterizing the deformation: the two complex constants $c,a$.
If the constant $c$ in the equation $zw =c$ is much bigger
than $\alpha'$, then we can trust the solution.
Notice that in this case, as opposed to the case discused in
sections 2 and 3,  we do not need to consider a large number of
branes to trust the solution, we can just tune a parameter in the
solution. Obviously, in the case of M-branes the condition
is that $c \gg l^2_{11}$.

It is an amusing exercise to see how this description
arises from the point of view of D-brane probes \refs{\mdgm,\dgauge}.
To that effect it is convenient to add two more dimensions
so that we are describing two D4 branes intersecting
along two dimensions. So we have a 4-brane along
$(z,t)$ and another 4-brane along $(w,t)$ where
$t$ is another complex coordinate describing
a two dimensional plane.
Then we can consider a D0 brane probe moving in this
geometry \juanandy .
Starting from the singular configuration we can write
a Lagrangian involving (0,4), (0,4') and (4,4') strings.
This action will have four real supersymmetries and can be
viewed as the dimensional reduction of a 4 dimensional
$N =1$ action. In fact the four dimensional action
describes a D3 brane probe in the presence of two orthogonal
D7 branes of the same $(p,q)$ type. We can therefore
use four dimensional $N=1$ superfield notation.
We will be interested in the superpotential
\eqn\superp{
S_{pot} = \int dt \d^2\theta \{
z_{00} \phi_{04}^+ \phi_{04}^- +
\omega_{00} \phi^+_{04'} \phi^-_{04'} +
\phi_{44'}^+ \phi_{04'}^+ \phi_{04}^- +
\phi_{44'}^- \phi_{04'}^- \phi_{04}^+ \}
+ c.c.
}
where $z_{00} , ~w_{00}$ are the chiral superfields  associated
to the position of the zero brane in the $z$ and $w$ plane.
$\phi^{\pm}_{04} $ are two chiral superfields associated
to the (0,4) strings and $\phi_{04'}^\pm $ are associated
to (0,4') strings. Finally $\phi_{44'}^\pm$ are associated
to (4,4') strings.
We now fix $\phi_{44'}^\pm ,~z_{00},~w_{00}$ and we diagonalize
the superpotential for the 04 and 04' fields.
After we do this we get new fields with holomorphic
mass parameters
\eqn\mass{
m_\pm = { z_{00} + w_{00} \over 2 } \pm \sqrt{
\left({ z_{00} + w_{00} \over 2 } \right)^2 -
\phi_{44'}^+ \phi^-_{44'} }
}
If we fix the value of the 44' strings we find that there
are some values of $z_{00}, w_{00}$ for which we have a
massless mode. Physically this means  that the
0-brane probe is touching the fourbrane.
We see that $m_\pm =0$ implies that $z_{00} w_{00} = \phi_{44'}^+
\phi^-_{44'}$, this is the same equation as we found before
for the shape of the single (in this case) fourbrane.
We see that exciting the 44' strings corresponds
to a complex deformation (some ``blow up'') of the
two intersecting 4 branes into a smooth configuration.

Another amusing exercise is to take one of the transverse fields to be
compact, $X^4 \sim X^4 + 2 \pi $ to be specific. In this case
$w = \log z$ is a solution. This solution
represents a membrane wrapped around the $\hat 4$ direction,
lying along $\hat 5$ and ending on a two brane oriented along 23.
This is how a fundamental
string ends on a D2 brane from the point of view of M-theory
\aharony . This is vortex solution since the field $X^4$
winds once when we go once around the origin.
This type of configuration (with some more branes) was used
to solve N=2 gauge theories in four dimensions in \wittenntwo .

These  results can be used to  count the entropy of black holes \ascv .
Suppose that we consider a system to $Q_3$ threebranes oriented
along $789$, $Q'_3$ D-threebranes oriented along $569$. Then
we can say that the $z$ plane is associated to $56$ and
the $w$ plane to $78$. We can see that we have
$Q_3Q_3'$ intersection points. At each of those points
we can make the deformation described above. This introduces
$4 Q_3 Q'_3 $ bosonic parameters, and an equal number of fermionic ones.
Which then give a CFT with central charge $c=6$ along the direction 9,
producing the right black hole entropy \ascv .
So the black hole entropy comes from deforming the D3-brane ``foam''
that results after ``blowing up'' each of the intersection
points. This is similar to saying that the entropy comes from 44' strings
\cama. Similarly, if we replace the $T^4$ along 5678 with a K3 we then
the threebranes will be some holomorphic two-surface on K3 with
$c_{IJ } p^I p^J/2 $ intersection points ($p^I$ are the vectors
specifying the charge configuration) and the counting
of the entropy proceedes in the same way \ascv .

%--------+---------+---------+---------+---------+---------+---------+

\newsec{Conclusions}

We have found some solutions of the \bi\ action describing
strings ending on D-branes, or M2-branes ending on M5-branes.
It would be interesting to see if one can pursue a
bit further this idea that lower-dimensional branes
arise from higher-dimensional ones.
We also found solutions describing a string going between
a brane and an anti-brane. We studied the propagation of
small perturbations on these solutions and found that the
cross sections had the properties required to understand
Polchinski's picture of D-branes as boundaries of open strings.
We also see that when we have  a brane-antibrane pair we
can, in some sense, interpret electric charges as some
kind of microscopic ``wormholes'' going from one brane to
the other in a purely geometric fashion.
This realizes Wheeler's old idea of `charge without charge' (the
existence of both electric and magnetic charges provides
the charge quantization condition).

%--------+---------+---------+---------+---------+---------+---------+

\bigbreak\bigskip\bigskip\centerline{{\bf Acknowledgements}}\nobreak

We would like to thank G. Horowitz for pointing out problems in an
earlier version of this paper related to
our interpretation of the Euclidean
bounce solution as a decay mode of a brane-antibrane pair.

We would like to thank M. Douglas, D. Lowe and J. Schwarz for
interesting discussions and for informing us about their work in this
area \dls. Also, after completing this work, we learned of a study
of related matters by Gibbons et al \gibbons. We also thank M. Kroyter
for pointing out a couple of typos. 

C.G.C.{} and J. M. were  supported in part by DOE grants DE-{}FG02-91ER40671
and DE-FG02-96ER40559 respectively.

\vfill\eject

%--------+---------+---------+---------+---------+---------+---------+

%\appendix{A}{ }

\vfill\eject
\listrefs
\bye